\documentclass[twocolumn,showpacs,pra,aps]{revtex4}
\usepackage{graphicx}

\begin{document}

\title{Classical Demons and Quantum Angels:\\On 't Hooft's deterministic Quantum Mechanics}

\author{Antoine Suarez}
\address{Center for Quantum Philosophy, P.O. Box 304, CH-8044 Zurich, Switzerland\\
suarez@leman.ch, www.quantumphil.org}

\date{May 27, 2007}

\begin{abstract}

It is argued that 't Hooft's deterministic program does not disenchant the quantum world but rather inspires the incantation of the classical one.\\

\footnotesize\emph{Key words}: Free will, quantum entanglement, before-before experiment, non-temporal causality, world's end, brain, consciousness, sleep, quantum homeostasis.

\end{abstract}


\maketitle

\section {Introduction}
In a recent paper Gerard 't Hooft proposes to substitute the ``free will postulate in Quantum Mechanics'', with a condition he calls ``unconstrained initial state''. The paper develops to a great extent a philosophical view about ``free will'':``the 'unconstrained initial state' condition has consequences similar to `free will', but does not clash with determinism [...]. The dismissal of the usual `free will' concept does not have any consequences for our views and interpretations of human activities in daily life, and the way our minds function'' \cite{hooft07}. The motivation for reintroducing determinism into physics is threefold:

1) To get a ``deterministic quantum mechanics'' capable of describing gravitational forces.

2) To overcome the ``unsatisfactory'' statistical nature of the quantum mechanical predictions, since ``only theories that, under ideal circumstances, would predict single events with certainty should be considered acceptable.''

3) To avoid concern about the transition point between quantum mechanical possibilities and classical certainties deriving from the ``collapse of the wave function'': ``When and how does such a `collapse' take place?'' When and how does the choice of the measured outcome among various possible ones happen?

It is 't Hoofts's belief that it is time to face such questions concerning the deeper nature of Quantum Mechanics, and that attempts to answer such questions will lead to further insights.

Sharing completely 't Hoofts's concern about the necessity of asking ``a next generation of questions'' that may allow us to understand Quantum Mechanics at a deeper level, I discuss in the following his proposal. \emph{Sections II-IV} define the meaning of the ``free-will assumption in Quantum Mechanics'' arguing that the Quantum World requires causes acting freely from outside space-time. \emph{Section V} argues that it is worth to work on developing deterministic quantizing models
of Gravity, as 't Hooft suggests. \emph{Section VI } discusses 't Hoofts's ``unconstrained initial state condition'' concluding that it conflicts with logic, freedom and experiment. \emph{Section VII} shows that it is in principle possible to unify indeterministic Quantum Mechanics and a deterministic description of Gravity: Unity comes from the assumption of causes or agents acting into the world from outside space-time, and diversity from the assumption that these agents may operate as well in a deterministic or indeterministic way. \emph{Section VIII} argues that there is an \emph{objective} transition point between the quantum possibilities and the classical certainties, but one should define it (i.e. when and how the choice of the observed outcome among various observable ones happens) with relation to the capabilities of the \emph{human observer}. \emph{Section IX} argues that freedom, rights and creativity exclude any explanation of the brain using only observable causal chains, and it is in principle possible to integrate a quantum mechanical view into neuroscience. In particular it stresses the relevance  of consciousness and sleep for physics. Finally \emph{Section X} summarizes the conclusions.

\section {The assumption of free-will on the part of Nature in Quantum Mechanics}
Laplace gave a good characterization of Determinism by means of his famous ``demon'':

``We may regard the present state of the universe as the effect of its past and the cause of its future. An intellect which at a certain moment would know all forces that set nature in motion, and all positions of all items of which nature is composed, if this intellect were also vast enough to submit these data to analysis, it would embrace in a single formula the movements of the greatest bodies of the universe and those of the tiniest atom; for such an intellect nothing would be uncertain and the future just like the past would be present before its eyes.''\cite{lap}

By contrast, according to Quantum Mechanics, Nature makes  choices that are \emph{not exclusively} determined by the past. This assumption originates from observations that can be summarized with the experiment sketched in Figure 1. When one works with sufficiently weak intensity of light, then only one of the two detectors clicks: either D($+$) or D($-$) (photoelectric effect). Nevertheless, for calculating the counting rates of each detector one must take into account information about the two paths leading from the laser source to the detector (interference effect). Both effects, single clicks and interferences, appear as well in experiments with particles like electrons or neutrons.

\begin{figure}[t]
\includegraphics[width=80 mm]{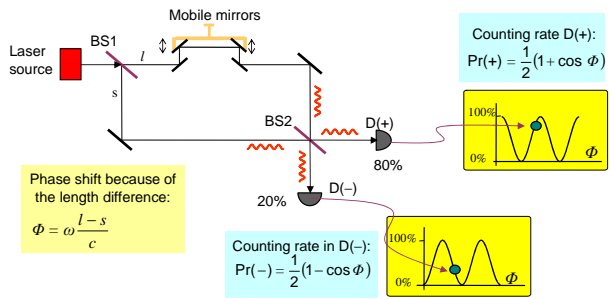}
\caption{Single particle interference are the entry (sort of ``Platform Nine-and-Three-Quarters'') to the Quantum World. Laser light of frequency $\omega$ emitted by the source is either transmitted or reflected at each of the beam-splitters (half-silvered mirrors) BS1 and BS2; therefore the light can reach the detectors D(+) and D($-$) by the paths $l$ and $s$; the path-length $l$ can be changed by the experimenter. With sufficiently weak intensity of light, only one of the two detectors clicks: either D($+$) or D($-$) (photoelectric effect). Nevertheless, Nature calculates the counting rates of each detector Pr(+) and Pr($-$) taking account of the length of the two paths $l$ and $s$ (interference effect).}
\label{fig1}
\end{figure}

Experiments, in which in each single run only one of two detectors clicks, and for a large number of runs the clicks are statistically distributed between the two detectors according to interference mathematics, are the entry to the Quantum World, just like Platform Nine-and-Three-Quarters is the entry to the Wizard World in Harry Potter. Indeed, according to the Copenhagen interpretation (mainly proposed by N. Bohr, W. Heisenberg, M. Born and P. Jordan) such experiments have three main implications:

1) Which detector clicks is decided by a choice (on the part of Nature) when the information about the two paths reaches the detectors.

2) To this aim, there is an agreement or exchange of information between D($+$) and D($-$), no matter how far away from each other these detectors are.

3) There is an indeterminacy about the time a particle is emitted from the source, which is inversely proportional to the frequency-band-width of the emitted light (Heisenberg's uncertainty principle).

According to 1), which detector clicks in a single run in the experiment of Figure 1 is not only unpredictable for us because we don't yet know the formula connecting the past with the present and the present with the future, but it is unpredictable in principle because such a formula doesn't exist at all. This is the quantum mechanical assumption of \emph{free-will or choice on the part of Nature}. The concept of ``free-will'' means here that Nature's decision about which detector clicks, though it has some roots in the past, is not \emph{completely} determined by the past, i.e. it has \emph{not all} of its roots in the past. In particular, the \emph{choice on the part of Nature} excludes the ``realistic'' view that the outcomes are determined by well defined properties the particles carry when they leave the source.

According to 2) in Nature there are connections happening faster than light and without propagation of energy. This is the quantum mechanical nonlocality assumption, which is implicitly contained in the so-called \emph{collapse of the wave function}, and led to the well known EPR controversy referred to in the next section.

Einstein was highly disappointed about these two features of Quantum Mechanics, and expressed his concern with comments like: ''God doesn't play dice'' and ``spooky action at a distance''.

Regarding 3), I would like to stress that Heisenberg's uncertainty principle should not be considered the base of Quantum Mechanics but a consequence of the more basic principles 1) and 2). It builds a guarantee for the self-consistency of Nature, i.e. for harmonizing the quantum mechanical phenomena with the classical concept of spacetime: Interferences are possible even if the velocity of light does not depend on the length of the path, i.e., light travels with equal velocity path $l$ and path $s$ in the experiment of Figure 1.

\section {Free-will on the part of the physicist and quantum nonlocality}

In the experiment of Figure 1 one can escape connections faster than light if one assumes that the choice happens at the beam splitter rather than the detectors. This view was supported by the interpretation Louis de Broglie proposed at the time of the Solvay Congress (1927). De Broglie's interpretation implies that the particle travels always one of the two possible paths, and an ``empty wave'' or ``pilot wave'' travels the other path. The ``empty wave'' is a pure information wave that guides the particle when both meet at a beam-splitter and makes interferences possible. The ``empty wave'' does not carry any energy and is completely unobservable. De Broglie's model may be considered local deterministic, in the sense that the outcomes are completely determined by information within the past light-cone. It may also be considered ``realistic'' in the sense that the particles are supposed to share always a definite position. Nevertheless, the calculation of the outcomes requires unobservable information (``empty wave'').

Notice that according to de Broglie's picture, the particle always travels a well-defined path. By contrast, according to Copenhagen the particle actually does not travel any path: the particle becomes real only at the detection, and the single detections (the ``clicks'') are distributed between the rival detectors according to wave mathematics, i.e. calculations taking account of all possible ``classical'' paths from the source to the detectors. The assumption that the particle makes a ``delayed choice'' between traveling one or two paths is a rather confused mix of de Broglie and Copenhagen.

The assumption that the choice happens at the beam-splitter avoiding connections faster than light may have helped to tranquilize the physicists for a time. But in 1935, Einstein counterattacked and advanced the EPR argument \cite{epr}. The EPR paper basically argued that in 2-particle experiments, as represented in Figure 2, the quantum mechanical view of an instantaneous ``reduction of the wave packet'' leads in anyway to correlations at a distance. Therefore, if one rejects nonlocal causality (``no real change can take place in the second system in consequence of anything that may be done on the first system'' \cite{epr}), one must accept that the correlations originate from ``real'' properties the particles carry when they leave the source. The EPR paper stressed very much the necessity of completing the wave function through additional observable variables.

\begin{figure}[h]
\includegraphics[width=80 mm]{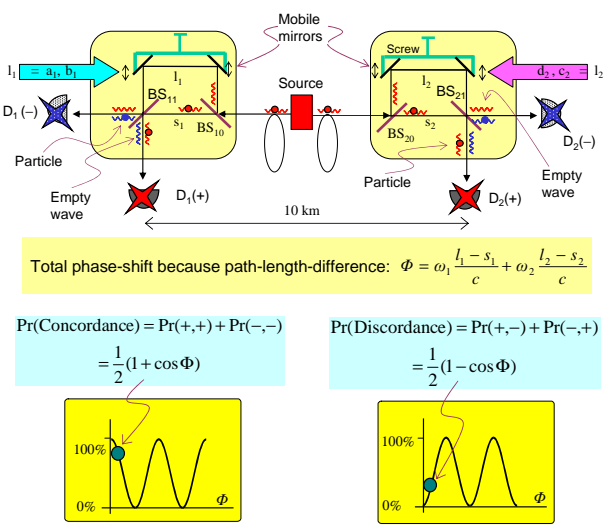}
\caption{In 2-particles Bell experiments local randomness and nonlocal order (correlations) appear inseparably united. The source emits photon pairs. Photon 1 (frequency $\omega_{1}$) enters the left interferometer through the beam-splitter BS$_{10}$ and gets detected after leaving the beam-splitter BS$_{11}$, and photon 2 (frequency $\omega_{2}$) enters the right interferometer through the beam-splitter BS$_{20}$ and gets detected after leaving the beam-splitter BS$_{21}$. The detectors are denoted D$_{1}(\varrho)$ and D$_{2}(\sigma)$ ($\varrho,\sigma\in\{+,-\}$). Each interferometer consists in a long arm of length $l_{i}$, and a short one of length $s_{i}$, $i\in\{1,2\}$. Frequency bandwidths and path alignments are chosen so that only the coincidence detections corresponding to the path pairs: $(s_{1},s_{2})$ and $(l_{1},l_{2})$ contribute constructively to the correlations, where $(s_{1},s_{2})$ denotes the pair of the two short arms, and $(l_{1},l_{2})$ the pair of the two long arms. Depending on the value of the phase $\Phi$, the quantity $Pr(\rho, \rho)$ oscillates between the situation of perfect correlation ($Pr(\rho, \rho)=1$), and that of perfect anticorrelation ($Pr(\rho, \rho)=0$). However, $Pr(\rho)$ for photon 1 alone is always 50\%, and $Pr(\rho)$ for photon 2 alone is 50\% too. Bell experiments, using two different values of $l_{1}$ and two different values of $l_{2}$, demonstrate that the correlations violate the locality criteria (Bell's inequalities).}
\label{fig1}
\end{figure}

David Bohm extended in 1952 de Broglie's idea of the ``pilot wave'' to experiments with several particles, and proposed a picture of Quantum Mechanics which aimed to give ``a precise, rational, and objective description of individual systems'' \cite{dbbh}. The particles always have a well defined position, in accord with Einstein's requirement (Figure 2). However, in the EPR situation, Bohm's formulation of Quantum Mechanics implies that the outcome at one beam-splitter precedes in time, and causes faster than light the outcome at the other beam splitter by means of a so-called ``quantum potential''. Additionally, Bohm's objective description can no longer be considered completely ``realistic'' since in experiments involving entangled polarized photon pairs neither of the two photons carries a definite polarization when it leaves the source (see \cite{grö} and references therein).

On the one hand, Bohm's picture implicitly contains the idea that quantum correlations cannot originate from the particles' local properties alone. As John Bell stressed: ``This picture [...] has the very surprising feature: the consequences of events at one place propagate to other places faster than light. This happens in a way that we cannot use for signaling. Nevertheless it is a gross violation of relativistic causality.''(\cite{jb87}, p. 171).

On the other hand, Bohm's theory can be considered an attempt to cast nonlocality into a time-ordered causal scheme. Bohm gives up the relativity of time and uses an \emph{undefined} preferred frame or universal clock: the value occurring later (the effect) in this preferred frame depends on the value occurring before (the cause).

John Bell took the next step and showed that: 1) local properties cannot generate correlations going beyond a certain degree, and 2) Quantum Mechanics goes beyond this degree: ``Bohm of course was well aware of these features [consequences propagating faster than light] of his scheme, and has given them much attention. However, it must be stressed that, to the present writer's [John Bell's] knowledge, there is no proof that any hidden variable account of quantum mechanics must have this extraordinary character. [Footnote: Since the completion of this paper such a proof has been found (J.S. Bell, Physics 1, 195 (1965)]''(\cite{jb87}, p. 11).

In more technical terms, John Bell showed that if one only admits relativistic local causality (causal links with $v\leq c$), correlations occurring in space-like separated regions should fulfill clear locality conditions (``Bell's inequalities'') \cite{jb64, jb87}.

Quantum Mechanics predicts correlated outcomes in space-like separated regions for experiments using 2-particle entangled states (Figure 2). Suppose one of the measurements produces the value $\rho$ ($\rho\in\{+,-\}$), and the other the value $\sigma$ ($\sigma\in\{+,-\}$). According to Quantum Mechanics the probability $Pr(\rho, \sigma)$ of getting the joint outcome $(\rho,\sigma)$ depends on the choice of the phase parameters characterizing the paths or channels uniting the source and the detectors; depending on the value of the phases, the quantity $Pr(\rho, \rho)$ oscillates between the situation of perfect correlation ($Pr(\rho, \rho)=1$), and that of perfect anticorrelation ($Pr(\rho, \rho)=0$). However, $Pr(\rho)$ for photon 1 alone is 50\%, and $Pr(\sigma)$ for photon 2 alone is 50\% too.

According to Einstein's local view of Relativity one has to exclude \emph{any} link faster than light and, therefore, any direct connection between space-like separated regions. This means that the quantum correlations could not arise from a direct link between the two measurements, but should originate from some common cause in the absolute past of both measurements: the correlations are pre-determined through certain hidden programs the photons carry when they leave the source. It is as if the photons would behave like genetical twins, and therefore the correlations they may produce should obey the Bell's inequalities.

However Bell proved that in experiments using measurements with different settings (for instance two different values of $l_{1}$ and two different values of $l_{2}$ in Figure 2) the correlations predicted by Quantum Mechanics violate the Bell's inequalities (Bell's theorem) \cite{jb64, jb87}.

Bell experiments conducted in the past two decades, in spite of their loopholes, suggest a violation of local causality:
statistical correlations are found in space-like separated
detections; violation of Bell's inequalities ensure that these
correlations are not pre-determined by some common cause in the past \cite{exp}. Nature seems to behave nonlocally, and Quantum Mechanics predicts well the observed distributions.

Nevertheless, even if the quantum nonlocality violates Einstein's view that nothing in Nature goes faster than light, it does not lead to any conflict with Relativity in the following sense: The quantum mechanical formalism prevents us from using the unobservable ``Bell connections'' for phoning or teleporting faster-than-light \cite{eb78}. It could not be otherwise since the Michelson-Morley experiment is an interference experiment, and therefore in fact a quantum experiment: there cannot be any contradiction between the consequences of this experiment and Quantum Mechanics. If Einstein felt a contradiction, it was because he extended the validity of Relativity beyond the limits permitted by the experimental data.

It is important to stress that a crucial assumption in Bell's theorem is \emph{the free-will of the physicist}. This means for the experiment of Figure 2 that in choosing the values of the path-lengths \emph{l} the physicists are not \emph{completely} determined by the past. Otherwise one could assume a sort of conspiracy in Nature guiding the physicist to set lengths fitting to the properties the particles carry in order to produce the correlations.

Another important feature of quantum entanglement is the unity of randomness and order. Bell experiments show that the quantum distributions do not result automatically from the real properties the particles carry and the real settings they meet, but require decisions taken actually place when measurement happens. Recent experiments with pairs of photons show that the nonlocal correlations happening in Nature and in Quantum Mechanics require even to give up the assumption that each single particle carries a definite polarization when it leaves the source. In other words, it is necessary to accept that the polarization Nature uses for producing the nonlocal correlations results from some arbitrary choice done at the moment of measurement \cite{grö}. However, in phenomena involving quantum entanglement there is not only randomness, but randomness and order together: An event A is nonlocally correlated according to statistical rules to another event B, and additionally A is correlated to B, and not to any other events near B. Behind results which locally look random, there is a nonlocal order.

\section {Behind the quantum phenomena there are free and intelligent causes acting from outside space-time}

Bohm's theory actually involves two different assumptions: 1) There are influences faster than light. 2) The event at one of the beam-splitters is the cause of the event at the other beam-splitter.

As a matter of fact, the conventional Bell experiments tackled only issue 1). They did not test whether or not entanglement depends on the timing of the measurements. Actually John Bell himself explained things on the basis of assumption 2) (and still today one proposes models invoking this assumption \cite{aa}). However, Quantum Mechanics assumes effectively (although implicitly) that the particles stay nonlocal correlated independent of any timing. Therefore, also taking nonlocal influences for granted after Bell experiments, experiments testing the timing-independence of the quantum probabilities were surely of interest.

In this respect it is worth to quote Bohr's answer (1935) to the EPR paper: ``It is true, that we have freely made use of such words as 'before' and 'after' implying time-relationships [...]. As soon as we attempt a more accurate time description of quantum phenomena, we meet with well-known new paradoxes, for the elucidation of which further features of the interaction between the objects and the measuring instruments must be taken into account. In fact, in such phenomena we have no longer to do with experimental arrangements consisting of apparatus essentially at rest relative to one another, but with arrangements containing moving parts, -like shutters before the slits of the diaphragms,-controlled by mechanisms serving as clocks.'' \cite{nb}

These words reveal that Bohr somewhat considered the possibility of experiments with moving devices, and time measurements by different clocks. A definite proposal for experiments testing this idea first came in 1997 \cite{asvs97}.

In planing such experiments the first problem one must fix is determining which clocks matter for the  time measuring. If one tries to cast nonlocal causality into only one preferred frame as Bohm does, it is not more reasonable to connect a ``cause'' event to an ``effect'' event in that frame rather than in some other frame. Effectively a single preferred frame (``quantum ether'') is ``experimentally indistinguishable" \cite{jb64}, and in fact Bohm's theory does not contain any clue about how to determine such a frame. The predictions would remain the same if one assumes that the preferred frame is a virtual entity changing from experiment to experiment. One is tempted to think that Bohm introduces ``absolute time'' just because he wishes to justify a causal description, but in the end, an untraceable ``quantum ether'' is essentially the same as deciding arbitrarily which event depends on which one. For all purposes Bohmian Mechanics can be considered a causal description but not a temporal one, in the sense of a time ordering that can be measured by a real clock, and to date it has not lead to experiments capable of distinguishing it from Quantum Mechanics.

Since the outcomes are supposed to become determined at the beam-splitters, and one cannot associate an inertial frame to the photons, it is natural to assume that the time at which the decision at one beam-splitter occurs is measured according to the clock defined by the inertial frame of this beam-splitter. The fact that all the reflected photons show a Doppler-shift  determined by the velocity of the device speaks also clearly in favor of this view: assumed the choice of the outcome occurs at the beam-splitter, it is the velocity of the beam-splitter which unambiguously defines the relevant inertial frame for determining the time of the outcome. Accordingly, the decision about the output port by which a photon leaves a beam-splitter takes account of all the local and nonlocal information available within the inertial frame of this device, at the instant the particle strikes it. Within each beam-splitters' frame the causal links always follow a well defined chronology. This was the basic assumption of the proposal to test whether nonlocal influences are compatible with time-orderings defined by real clocks \cite{asvs97, as00.1}.

Notice that as long as one believes (according to Einstein) that there are no space-like influences, the fundamental temporal notion could not be other than proper time along a time-like trajectory. But since Bell experiments did reveal a world consisting in nonlocal connected events, the ``reasonable'' position in the very spirit of relativity is to assume time-ordered causality, and describe the nonlocal links using lines of simultaneity to distinguish between ``before'' and ``after''. Indeed, such a description is very well possible in conventional Bell experiments, in which all apparatuses are standing still in a laboratory frame. Since the emission time of the photons is not exactly the same, and the fibers guiding the photons from the source to the measuring devices do not have exactly the same length, according to the clock defined by the laboratory's inertial frame, one of the measurements always takes place before the other, and the particle arriving later can be considered to take account of the outcome of the one arriving before. In experiments with all measuring devices at rest, it is possible to explain quantum correlations through time-ordered (nonlocal) causality.

However, in experiments with moving apparatuses, several relevant frames are involved: different clocks watch the arrival times, and what is ``after'' according to the laboratory clock may become ``before'' according to one moving clock. Consider an experiment in which the beam-splitters are in motion in such a way that each of them, in its own reference frame, is first to select the output of the photons (\emph{before-before} timing). Then, each outcome will become independent of the other, and the nonlocal correlations should disappear \cite{asvs97,as00.1}.

As said, Quantum Mechanics assumes that the particles stay nonlocal correlated independently of any timing, even in such a \emph{before-before} situation. This means that \emph{before-before} experiments are capable of acting as standard of \emph{time-ordered} nonlocality (much as Bell's experiments act as standard of locality): if timing-independent Quantum Mechanics prevails, nonlocality cannot be imbedded in a relativistic chronology; if Quantum Mechanics fails, there is a time ordering behind the nonlocal correlations, and proper time along a time-like trajectory is not the only temporal notion.

Acousto-optic modulators have made it possible to perform
experiments with  beam-splitters moving at 2500 m/s and realize the \emph{before-before} situation \cite{as00.1}. The setup is represented in Figure 3. The results obtained in June 2001 uphold the predictions of Quantum Mechanics \cite{szsg}.

\begin{figure}[t]
\includegraphics[width=80 mm]{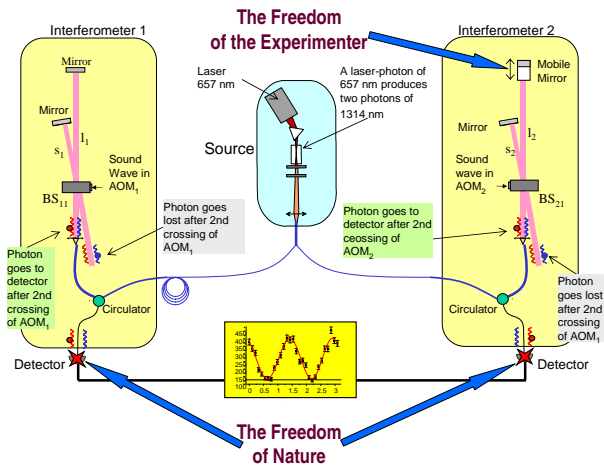}
\caption{The ``before-before'' experiment \cite{as00.1, szsg}: As beam-splitters one uses sound waves produced in acousto-optic modulators (AOM), and propagating at a velocity of 2500 m/s.  Thus, the beam-splitters are in motion in such a way that each of them, in its own reference frame, is first to select the output of the photons. Then, each outcome becomes independent of the other, and the nonlocal correlations should disappear. However experiment shows that the nonlocal correlations appear independently of the \emph{before-before} time ordering. This means that the order of the laws of Nature is not established by time. Behind the quantum phenomena there are free and intelligent causes acting from outside spacetime.}
\label{fig1}
\end{figure}

The influences allowing us to phone between two separated regions follow time-like trajectories, and can consistently be described in terms of ``before" and ``after" by means of real clocks; Einstein's world contains only such local causal links. The entanglement bringing about nonlocal correlations is not sensitive to space and time, and cannot be described in terms of ``before'' and ``after'' by means of any set of real clocks. The notion of time makes sense only in Einstein's world, i.e. along time-like trajectories.

The results of experiments with moving beam-splitters mean that the quantum correlations are caused regardless of any relativistic chronology: entangled photons run afoul of the relativity of time, Einstein's frames have no effect on ``spooky action'', even though we cannot use this fact to establish an absolute time. So what Quantum Mechanics actually implies is that in case of space-like separated measurements the connection the correlations reveal does not correspond to any real time ordering and, consequently, is not tied to any experimentally distinguishable frame. It is not possible, even in principle, to distinguish which measurement is the independent and which the dependent one. In this sense, the experiment rejects models invoking a particular frame of reference and stating that measurement on particle 1 happens first in time, and that its result immediately affects the state of particle 2 \cite{aa}. The ``before-before'' experiment definitely rules out the view that one of the events at the beam-splitters is the cause of the other one.

Suppose an intellect could act non-locally and would like to bring about Bell-correlations. this intellect would first choose one event assigning randomly a value (either $+$ or $-$) to it, and subsequently would assign a value depending upon the first, to the second specific event. Suppose these operations occur without the flow of time; as Quantum Mechanics seems to mean, and the experimental results confirm, this is the way things happen in Nature \cite{as03}.

It is also worth to stress that according to Quantum Mechanics the statistical distribution corresponding to a determined phase \emph{$\Phi$}, say 80 \% counts in detector D($+$) and 20\% in detector D($-$) in the experiment of Figure 1, appears even if after every single detection one would replace the two detectors through two new ones. This means that the information of what has already been counted remains stored in some memory without need of any material carrier.

In conclusion, the experiments testing quantum entanglement rule out the belief that physical causality necessarily relies on observable signals, and that an observable event (the effect) always originates from another observable event (the cause) occurring before in time. This means that the quantum correlations have roots outside the space-time and, in this sense, originate from a free and intelligent agent. One is led to accept `` the two freedoms'': the freedom of the experimenter and the freedom of Nature \cite{az}, and to see quantum randomness as a particular expression of free will. Actually I think that the experiments with moving devices described above are a stronger proof of the free-will theorem than that proposed in \cite{coko}: If I assume to be \emph{someone} rather than \emph{nothing}, if I assume to share intellect and free will, if I claim in particular that the author of this article is not exclusively the Big-Bang, I have also to accept that there is some intellect behind the quantum phenomena rather than assuming that there is \emph{nothing}.

All these results support the view that the quantum distributions originate from an enormous amount of calculations  and decisions unobservable intellects (Nature) do taking account of the experimental setup. ``Wave functions'' exist and evolve in such mighty minds, which do not require a brain to know and act. By contrast-analogy to the \emph{classical demons} (Laplace's and Maxwell's ones) I would like to propose naming the non-neuronal intellects performing the quantum mechanical distributions as \emph{quantum angels}.

\section {\emph{Indeterministic} Quantum Mechanics may be inappropriate for explaining Gravity ...}

Gerard 't Hooft forewarns very definitely that indeterministic Quantum mechanics ``fails when gravitational forces come into play'', and ``superstring theory does not appear to give us hints as to where such solutions will come'' \cite{hooft07}. It is this failure that leads him to the belief that explaining gravity requires \emph{deterministic} Quantum Mechanics.

On the one hand, such a warning does not come from lack of knowledge about how efficient indeterministic Quantum Mechanics can work: 't Hooft has substantially contributed to establish the Standard Model, which ``describes in meticulous detail how all known constituents of matter behave. It is an entirely quantum mechanical description, and one can say that Quantum Mechanics has emerged as an absolutely indispensable framework for this description.'' On the other hand, the warning does not reflect lack of effort or expertise in the field of Quantum Gravity: 't Hooft himself is working since about 20 years to the unification of Quantum Mechanics and Gravity, and has achieved results like his holographic principle, which is a major contribution to the field. In view of 't Hooft's CV, one should take his warning seriously.

One understands that the success of gauges theories and the Standard Model invites very much to quantize gravitational fields as one quantizes electromagnetic ones. However, the experiments founding Quantum Mechanics, like that of Figure 1 and 2, rely on the possibility of defining at least two different optical paths leading from the same source to the same detector. And such a possibility relies on the fact that light can be screened, and guided.

An indeterministic quantum theory of gravity must also basically assume that gravitational waves can go from one same source to one same detector by two different ``paths''. The impossibility of screening and guiding gravitation (``gravity couples to everything'') leads to conceive these ``paths'' as different geometries rather than worldlines. In any case, a quantum mechanical theory of gravity means that it is possible to produce interferences with gravitational waves coming from one single source. Therefore, any model trying to apply indeterministic Quantum Mechanics to gravitational forces should include a proposal for such interference experiments. Without this it is surely worth continuing to work on developing deterministic quantizing models of Gravity. But, even taken for granted that explaining gravity requires a \emph{deterministic} quantizing procedure, this is not a sufficient condition to assume that one can explain the quantum phenomena, and in particular quantum entanglement, using exclusively a deterministic procedure.

\section {... but 't Hooft's ``unconstrained initial state condition'' clashes with logic, freedom, and experiment}

Gerard 't Hooft defines its ``unconstrained initial state condition'' as ``the possibility to freely choose our initial states'': If we would have been deprived of the possibility to freely choose our initial states, we would not know whether our model makes sense at all: ``we must demand that our model gives credible scenarios for a universe \emph{for any choice of the initial conditions!}.'' \cite{hooft07}

Gerard 't Hooft quotes the following statement of the mathematician Conway: ``We have to believe in free will to do anything; I believe I am free to drink this cup of coffee, or throw it across the room. I believe I am free in choosing to have this conversation.'' \cite{co}. And then he comments: ``But of course Conway should know that, in spite of his apparent freedom to throw his coffee across the room or not, whatever he actually does is determined by laws of physics, not by some mysterious, unspecified, 'free will'. This, at least, is the real implication of the assumption of determinism. He is free to choose what to do, but this does not mean that his decision would have no roots in the past.'' \cite{hooft07}

Nevertheless, 't Hooft says, the universe, because its complexity, for all practical purposes remains for us as unpredictable as it is in the quantum mechanical description: ``even the best conceivable computer, cannot compute ahead of time what Mr. Conway will do, simply because Nature does her own calculations much faster than any man-made construction, made out of parts existing in Nature, can ever do. There is no need to demand for more free will than that.''\cite{hooft07}

Gerard 't Hooft rejects the Copenhagen interpretation because it ``carries a certain amount of agnosticism'': ``We will never be able to determine what actually happened during a physical experiment, and it is asserted that a deterministic theory is impossible. It is this agnosticism that we disagree with.'' \cite{hooft99}. Implementing the quantum mechanical view in string theory ``one discovers that we cannot rigorously define what quantum mechanical amplitudes are, what it means when it is claimed that `the universe will collapse with such-and-such probability', what and where the observers are, what they are made of''\cite{hooft99}.

The very basis of 't Hoofts's model seems to be his conception of causality as exclusively temporal causality: ``The notion of time has to be introduced if only to distinguish \emph{cause} from \emph{effect}: cause must always precede effect. [...] Since laws of Nature tend to generate extremely complex behavior, their effects will surely depend on the order at which they are applied, and our notion of time will establish that order.''

\subsection{Logical oddities}

Assuming, as 't Hooft states, that there is ``only one kind of logic'', I dare to say that his ``unconstrained initial state condition'' bears some logical oddities:

If the notion of time establishes the order of the laws of Nature and is the essential ingredient for defining causality, one wonders how there can be an ``initial state'' at all. If one assumes an initial state in time, then one accepts that the cause of this initial state is outside time, and therefore time cannot be the essential ingredient of causality.

It seem also to us that Mr. Conway is well aware that he would not be creating his coffee, at the moment he  would ``throw his coffee across the room'', and therefore Conway is not claiming at all that ``his decision has \emph{no} roots in the past''. Accordingly, regarding Mr. Conway, it may likely be more fair to say that, in postulating free-will, what he claims is rather that his decision has not all roots in the past, i.e. is not completely determined by the past. Certainly, ``whatever he actually does is determined by laws of physics'', but for the time being it sounds a bit assuming to pretend that the ``quantum mechanical laws'' do not deserve the status of laws of physics. And if whatever Mr. Conway actually does, follows quantum mechanical laws, then he is perfectly coherent and rational in claiming that he is free, in doing what he does.

If I understand well 't Hooft's definition, he seems to assume that free choice in the sense of Quantum Mechanics does actually exist but only as the possibility the scientist has for choosing the initial conditions in order to develop a mathematical proof, so for instance to see whether some initial conditions would lead to a contradiction. In this sense he seems to assume that thinking requires freedom (a view Kant already shared, and I fully share too). Nevertheless, choosing some initial conditions in a proof means nothing other that setting certain neural assemblies in your brain, to the aim of studying how these assemblies evolve. But this is exactly the same as an experimenter does to the aim of changing the settings in the setup of Figure 2. If we accept that the choice of the initial conditions in a proof has not all roots in the past, then one should also assume that the choice of the experimental settings has not all roots in the past.

Additionally, even admitted that today models should be able to describe credible scenarios for any choice of the initial conditions, one may doubt that some physicist was present at the Big-Bang performing the very choice that actually and deterministically led to the present state of the world affairs. Let us assume it was God, as an uncaused cause, who performed this choice. But then one wonders why God's choice was so weird to produce two classes of scientists contradicting each other: those believing in free will, and those denying it. It seems more logical assuming that in the beginning God created also the quantum randomness, and thereby make it possible the two classes of scientists to arise.

In summary, 't Hooft's view seems in the end to reduce to a conspirational theory in which the initial conditions of the universe are so contrived that the scientist cannot help to set the experimental parameters according to the properties the  EPR pairs carry so that the correlations arise. For sure, in a deterministic theory, this 'conspiration' is difficult to object against. So finally, the whole question seems to to be: ``Either you are for freedom, or not''. And finally this is also a matter of free choice ...

\subsection{Incompatible with freedom}

The deterministic view bears problems also regarding freedom as the basis of law and social life:

Disposing of the usual free will concept, and stating that there is no need for more free will than that leading to the practical impossibility for a human being to predict the future, may have severe consequences ``for our views and interpretations of human activities in daily life, and the way our minds function'': If my deeds are determined by the initial conditions at the beginning of the universe, I can assume to act under command of these conditions and do not feel responsible at all for the harm I could do to others. But the conscience of such a responsibility, and not only the fear of possible penalties, is presumably essential for the functioning of any free society.

As Conway states, free will is a capability we daily use. If free will is an illusion, societies relying on this capability should go wrong. Interestingly, History shows rather the opposite: disposing of the principle that humans are free to choose what to do, brings up dreadful societies. In fact, inside the deterministic logic the following grave implication seems inescapable: when someone purporting determinism comes to political power, and has to take decisions concerning human activities in daily life, he will tend to identify himself with a determining ``superior might'' choosing the conditions for the evolving of society, and consider the others as beings without free will, who have no need to demand for more freedom than that of ``the uncertainty about their future''. The others, if they think also deterministically, will agree in performing the decisions of the ``superior might'' without feeling any responsibility for their deeds.

It is somewhat astonishing that great determinist thinkers like Spinoza and Laplace overlooked such obvious consequences. Kant was aware of them and wanted to keep the causality of freedom while declaring that it was incompatible with the causality of nature. Somewhat Kant proposed the contradiction as philosophical way of life. The reasonable attitude, I dare to say, had been to state: If thinking requires freedom, physics cannot be deterministic and some day it will be completed  through some indeterministic description. This is what happened when Quantum Mechanics arrived, and in particular the Copenhagen interpretation.

Therefore, it seems to me that this interpretation does not carry any agnosticism at all. As said, I think that a deterministic theory of gravity is possible, but not a deterministic theory of everything. Asserting that a deterministic theory (of everything) is impossible, Copenhagen renders freedom possible \cite{ssb}. Indeed the failure to predict the single event Copenhagen assumes, is a merit rather than a defect. Consider the brain of professor 't Hooft. I think it is impossible to make a theory capable of predicting which will be the precise title of the next e-Print this brain will produce for arXiv.org, and this not only for reasons of complexity, but since the title involves free decisions this brain has not yet made, i.e., decisions whose roots are \emph{not all} in the past. Would \emph{all} of 't Hooft's decisions be completely determined by the Big-Bang, then he would have no merit for his discoveries, and it is the  Big-Bang, and not the professor, who should have been rewarded with the Nobel Prize. If there is \emph{someone} behind the single events in the professor's brain, then quantum indeterminacy is a necessary ingredient for the way human brains function. Hopefully, since otherwise a day will come in which neuroscientists will be capable of putting us under the Imperius curse!

\subsection{In conflict with experiment}

Even more than from these logical and anthropological oddities, the main problem with 't Hooft's proposal for a \emph{deterministic} quantum theory of everything may come from the incompatibility of his causality view with the experiments referred to above:

Indeed, for the time being, it is not clear how one can from variables at the Planck scale, by means of mere local causality, reproduce quantum distributions that are not only nonlocal but absolute timing independent. In our view, as Classical Physics up to now, any future sharp \emph{local} description of reality will have to define its variables with relation to classical properties we can change like we wish \cite{as02}. One is then necessarily led to the Bell's locality criteria.

More important: As 't Hooft acknowledges, his deterministic model does not yet explain ``(destructive) interference''. But this means that the model has not yet entered the quantum realm. If you pretend to have a quantum mechanical description of the world you have to be able to explain experiments as those of Figure 1-3. If one wishes to develop an alternative theory to indeterministic Quantum Mechanics one should care from the very beginning that the theory is capable of explaining quantum mechanical interferences.

Finally, it may be wise to keep in mind that 't Hooft's is not the first attempting to turn back the wheel of history. Einstein tried to demolish quantum indeterminism, and Quantum Mechanics is still alive. More specifically, Bohm tried to develop a ``deterministic Quantum Theory'', and he was led to nonlocality! And taken nonlocality for granted, I myself tried to keep to temporal causality and demolish at least the quantum mechanical timing-independence. But timing-independence prevailed. Anyway 't Hooft's view of causality, in which the notion of time establishes the order, seems to conflict from the very beginning with the ``before-before'' experiment, which shows that Nature establishes order without time.

\section {Is it possible to unify indeterministic and deterministic quantizing procedures?}

The preceding discussion rises the question about whether it is possible to combine an indeterministic Quantum Mechanics capable of describing two paths interferences with a deterministic quantizing procedure describing Gravity.

At this point, it is useful remembering that the crucial reason for quantizing General Relativity is that it bears black hole singularities. At a singularity the deterministic laws of physics no longer apply. Nevertheless, all known singularities appearing as solutions of Einstein's equations are always screened by ``an event horizon'', which prevent us from observing the singularity. Thus, whatever an outside observer sees is determined by usual laws of physics and whatever happens beyond the horizon is totally irrelevant for the phenomenology of a black hole \cite{ho04}. Determinism still works for all observational purposes. But one is led to accept the existence of domains of physical reality in which one cannot even predict how space-time behaves by means of deterministic equations. This is exactly the opposite of what General Relativity is supposed to be able to do. By predicting black holes, the theory somewhat predicts its own collapse.

As first step to overcome this oddity, I propose to keep to Bohr, and acknowledge ``the necessity of a final renunciation of the classical idea of causality and a radical revision of our attitude towards the problem of physical reality'' \cite{nb}. In fact this means to generalize the insights of the ``before-before'' experiment, and give up definitely the idea that each observable event can be explained exclusively by observable causes in the past, \emph{also when describing the classical deterministic world.}

As already noted above, in his deterministic model 't Hooft himself actually assumes causality coming from outside the space-time. For if one assumes an initial state in time, then one necessarily accepts that the cause of this initial state is outside time.

A possible way to get rid of temporal observable causality when describing the classical world it may consist in conceiving it as an \emph{animation} performed by the same non-neuronal intellects who are behind the quantum phenomena. In this view the movements of trains or planets would really very much happen as they do in the pretty animations by professor 't Hooft in his homepage \cite{hohp}. Entities 't Hooft calls ``changeables'' or ``ghost operators'' would correspond to nothing other than the operations performed by such invisible intellects.

Again, 't Hooft himself seems to suggest the \emph{animation} picture when he states: ``Nature does her own calculations much faster than any man-made construction, made out of parts existing in Nature, can ever do'' \cite{hooft07}. If ``Nature calculates'', then there are minds calculating, and the calculation has to stop at a certain point. So for instance the orbit of the planet Mercury or the bubble trail of an electron in a bubble chamber result through a sort of plotting and erasing information pixels at some extremely tiny scala. Information disappears at a time and appears again the next unit of time.

According to this view the whole visible body is at once the source and the detector, it consists in units of information appearing and disappearing very quickly, just like the desktop picture on a computer screen. Between two units of time there is no space-time and no matter, but only something like a wave function in the mind of the invisible, non-neuronal intellects performing the animation.

The non-neuronal intellects responsible for the classical animation are the same as those calculating the quantum mechanical distributions. In quantum phenomena they combine both indeterministic and deterministic procedures: on the one hand, for each single event they decide the outcome at will, and calculate the distribution of the single events taking account of the information about the different paths between the source and the detectors according to the quantum mechanical rules; on the other hand they calculate the  evolution in time of the quantum statistical distributions  according to deterministic procedures like Schr\"{o}dinger's equation. In case of the classical phenomena they calculate all the way using deterministic procedures: the information content of a pixel P is completely determined by the content of the pixels in the past light-cone of P, according to deterministic equations of movement. Nature calculates time evolving quantum distributions and visible bodies according to similar deterministic equations.

In the \emph{animation} picture the concept of \emph{information} becomes fundamental for classical deterministic physics too. The information contained in a pixel P of the space-time is completely determined by the information contained in the pixels building the past light cone of P. But it does not make sense to say the pixels in the past light-cone of P are the cause of P, because between pixels there is no ``bridge'', no reality (no information) at all. Information pixels are conditions, not causes, even if the appearances may lead to the illusion that they are causes, as Conway's Game of Life illustrates pretty well \cite{jc}. In this sense Hume's critics of temporal causality speaks in favor of the \emph{animation} picture too. The ``reversibility of the physical laws'' is a mathematical illusion: behind any physical phenomenon there is a decision.

The view I am proposing, very much like the Copenhagen interpretation, implies that a physical description makes sense as far as its theorems are linked to available or possible observations. Therefore it does not make sense to ask what happens within a region from which in principle no information can reach a human observer. I call \emph{death horizon} the boundary around a black-hole beyond which the functioning of any human brain will irreversibly break down and the human observer can be declared clinic death. According to the size of the black-hole, the \emph{death horizon} can be larger, equal or smaller than the \emph{event horizon} (in case of a very large hole, a human observer could cross the event horizon, and write a master work before dying). I call \emph{end of the world} the smaller of the two horizons, \emph{death} or \emph{event horizon}: who travels into a black hole will arrive at a world's end. Because no information from beyond a \emph{world's end} can reach any human observer, beyond this \emph{end} any calculation of Nature necessarily stops.

This means that a black hole singularity has no physical reality. This also means means that any information falling into the hole must remain stored at a \emph{world's end}, much the same way as information about detections remains stored in some immaterial memory and not in the detectors (see the end of \emph{Section IV}). In this sense it is natural to assume that, if a black-hole evaporates, the resulting radiation should be capable of informing observers in the outside region about what falls into the hole and about new information eventually created between the event and the death horizon.

Regarding quantum computers Gerard 't Hooft dares to predict: ``In our view, it [decoherence] will be one of the essential obstacles that will forever stand in the way of constructing super powerful quantum computers, and unless mathematicians find deterministic algorithms that are much faster than the existing ones, factorization of numbers with millions of digits will not be possible ever.'' \cite{hooft99} Fully acknowledging the formidable challenge of decoherence, it seems to me that this time it is 't Hooft who exhibits a remarkable amount of agnosticism and a not very enlightened view. As far as decoherence does not lead to irreversibility (see Section VIII), I am confident that super powerful quantum computers will be possible, as electron and scanning tunneling microscopes are possible. I am confident also about the possibility of super accurate and powerful information technologies exploiting the deterministic procedures Nature uses for ``plotting'' and ``erasing'' pixels at the smallest scala.

If the calculations responsible for the classical phenomena are performed by the same intellects performing the quantum mechanical distributions, 't Hooft's program for deterministic methods quantizing gravity does not disenchant at all the Quantum Realm but rather suggest that one has to enchant the Classical one.

\section {The transition point between ``quantum mechanical possibilities and classical certainties''}

I discuss now motivation 3) in 't Hooft's proposal: to overcome the problem of the transition point from quantum to classical, the point ``at which observations turn into classical certainties''. The problem was brought into focus already by Bohr in his answer to the EPR paper: ``While, however, in classical physics the distinction between object and measuring agencies does not entail any difference in the character of the description of the phenomena concerned, its fundamental importance in quantum theory, as we have seen, has its root in the indispensable use of classical concepts in the interpretation of all proper measurements, even though the classical theories do not suffice in accounting for the new types of regularities with which we are concerned in atomic physics.'' \cite{nb}.

As far as one accepts that there are both, a quantum indeterministic and a classical deterministic realm, the question naturally arises: where do we draw the line between the two realms, or in Wheeler's wording, when does a process of amplification become irreversible and produce a registered phenomenon? \cite{jw}.

If our \emph{animation} view is correct, once an information is registered, for instance the first bubble of the bubble trail of an electron in a liquid hydrogen chamber, successive plots will deterministically follow in natural units of time, in the corresponding observer's frame. But when precisely does the wavefunction collapse and the first bubble appear? The same question can be put with relation to the detection of a photon by a photomultiplier: When exactly after the wave's arrival can we say the detection takes place?

In this context ``gravitationally induced decoherence'' \cite{rp} and ``spontaneous wave function collapse'' \cite{bg} have been invoked. These processes are also denoted ``objective reduction''(OR) or ``unconventional decoherence processes'' \cite{rp} because they dispose of the assumption that a human observer has to be actually present for a registration to take place. But it may also be profitable to consider a daily process that is generally considered to be irreversible in principle: I mean \emph{death}, already referred to above with relation to the black holes. The clinical definition of death includes explicitly the concept of irreversibility as it basically states that death occurs when the neural functions responsible for certain spontaneous movements irreversibly breakdown. In establishing death this way, we are assuming as obvious that our capacity of restoring neuronal dynamics (our repairing capability) is limited in principle, even if we don't yet know where this limitation comes from. In an analogous way one could assume that a process of amplification in a photomultiplier becomes irreversible in principle at a certain level, if as soon as this level is reached an operation beyond the human capabilities would be required to restore the photon's quantum state. When such a level is reached the detector clicks. Such a view combines the subjective and the objective interpretation of measurement: on the one hand no human observer has to be actually present in order a registration takes place, just the same as in the GRW ``spontaneous collapse'' \cite{bg}) or Penrose's ``objective reduction'' (OR) \cite{rp}; on the other hand one defines the 'collapse' or 'reduction' with relation to the capabilities of the human observer.

Though getting a sharp formulation of ``measurement'' may depend of understanding what ``irreversibility'' means in the definition of death, ``measurement'' means creation of information too. According to Anton Zeilinger a detection is an individual act of creation. When the ensemble of quantum mechanical possibilities (the wavefunction) breaks down, one of the various possible outcomes becomes reality. When detection happens, new information is put into the world. Therefore it seems plausible to assume that the size of the neuronal assemblies required for generating consciousness (see \emph{Section IX} below) establishes the minimal amount of energy the ``click'' requires.

The view proposed here (very much like Bohr's view in his answer to EPR) implies that there cannot be such a thing as a ``wave function of the whole universe'', which would also include all human observers. If there is no human observer outside the wavefunction, there is no wavefunction at all. The fundamental importance of the human observer in Quantum Mechanics is the obvious consequence of the fundamental importance of observation and evidence in science: No science without observation, and no observation without observer. What Quantum Information seems to tell us after all is that the world is a dialogue between mighty non-neuronal intellects and human ones. Only someone mad keeps speaking if there is nobody to hear at him. And Nature is not mad.

These arguments suggest that the assumption of a transition point between quantum and classical is not absurd, though ``the measurement problem'', i.e. where must we precisely draw the line between the two realms, remains a mystery still to elucidate in today's physics.

\section {Towards a physics of consciousness and sleep}

Anyone who accepts freedom must necessarily reject any explanation of the brain using only deterministic causality, be it in terms of genes, chemicals or environmental influences: anyone believing that as a person he can escape genetic programming and ``chose how to live his life'' should coherently exclude any explanation of the brain including only observable causal chains. Anyone who is for freedom and claims for his rights, necessarily assumes that his free will is somehow involved in governing certain movements of his human body. When I speak or write, I implicitly assume that I am governing the movements of my lips, hands and eyes through the spiritual powers of the soul, i.e. my free will and intellect.
The way we behave in daily life, the way we organize society through law, even the successful implementation of quantum mechanical methods in financial markets provide evidence supporting an explanation that combines both, deterministic and quantum causality.

Many features of my brain's physiology are susceptible of deterministic description in terms of observable causal chains (the metabolism involved in the arousal potentials triggering bodily movements, for instance, follows the usual physical conservation laws). By contrast, the choices guiding my spontaneous movements, for instance typing a particular key ('r', 'a', 'd', 'o', etc.) while writing this paper, they originate from unobservable spiritual agency, i.e. my \emph{self} or \emph{soul} \cite{js}: I am obviously their author and responsible for them, as Mr. Conway would be responsible for throwing his coffee across the room, and would eventually be ordered to clean the floor. This quantum philosophical view overcomes the dualistic view of the soul ``as something mental, divorced from the tangible grey matter''.

But at what level in the brain does the choice determining whether someone moves his right or left hand take place? We know today that this depends on the building of different transient neuronal assemblies. But which processes may be responsible for the difference? The neuronal assemblies (like the counts in different detectors in quantum experiments) are measurable. But the cause choosing between two rival neuronal assemblies, as the cause choosing between two rival detectors, may very well be unobservable. Our prediction is that the realization of one specific neuronal assembly among several possible ones cannot be explained exclusively through deterministic causality but depends on quantum effects.

For the time being, a big difficulty in tackling this issue by experimental means originates from the high inaccuracy of current measuring techniques: imaging techniques for instance are still too slow to capture the recruitment of ten million cells in less than a quarter of a second. Nonetheless understanding how alternative behavioral patterns (moving this way, instead of another way) originate in the brain is in our opinion crucial for neuroscience, and even inaccurate exploratory experiments may be beneficial to progress.

Besides mathematical undecidability (G\"{o}del and Turing theorems) and the impossibility of signaling faster than light, there is a third fundamental limitation defining the human condition: the impossibility of uninterrupted consciousness. The human mind cannot be \emph{continuously} conscious (i.e. aware of his own existence), the human will cannot act on purpose all the time. If \emph{consciousness} requires a certain size of neural assemblies in our \emph{cortex} \cite{sg}, such assemblies undoubtedly represent an important achievement of evolution. But I dare to state that evolution never will be able to produce a brain capable of achieving consciousness without having to pay for it with sleep.

A human person can be considered a unity of consciousness and sleep, a neuronal intellect. Our brain is a device combining both, meaningless spontaneous behavior and purposefully ordered one. An image, crude though it may be, can help to suggest a way how this could happen. Imagine a brain operates to a certain extent like a quantum interferometer (Figure 1). A single outcome, either '$+$' or '$-$', represent a bit of information, and a large series or string of outcomes (a bit string) build a piece of information. The physiological measurable parameters of the interferometer are fixed by a number of factors (genetic, epigenetic and environmental ones), and in particular by sensorial impressions. These parameters determine the statistical distribution of the outcomes for large series at a given time, say for instance: 53\% '$+$' and 47\% '$-$'. During certain periods of time the interferometer produces meaningful pieces of information (speech, text, musical composition, painting...), where the meaning is given by the order of the bits in the string. During other periods of time (while sleeping and even during many wake periods) the interferometer produces meaningless strings of bits. Hence, I fully support the view that ``the causal efficacy of mental effort is no illusion'', but I think one can have this without invoking the Quantum Zeno Effect \cite{ssb}. It seems more plausible to assume that the efficacy of consciousness originates from influencing the order of the bits in the string of outcomes without violating their statistical distribution for large series.

According to the \emph{Synaptic Homeostasis Hypothesis}, ``sleep is the price we have to pay for plasticity, and its goal is the homeostatic regulation of the total synaptic weight impinging on neurons'' \cite{tc}. As a complementary hypothesis, I propose \emph{Quantum Homeostasis}: the homeostatic regulation of the quantum statistical distribution of brain outcomes. In producing wilful and meaningful pieces of information (short series of bits), the statistical distribution of the brain outcomes may differ to some extent from the distribution the physiological parameters impose for large series of outcomes. Such a statistical deviation needs to be balanced by other periods (sleep, uncontrolled movements) restoring the large scale distribution. (Such a restoration, and in this sense Quantum Homeostasis, is already at work in any quantum random number generator). In particular I claim that REM sleep is important not only to the developing of the infants brain, but also to the homeostatic regulation of quantum distributions in each phase of life: REM sleep is the price we have to pay for rather short periods of intentional and meaningful behavior. In this perspective the unity of consciousness and sleep requires a world governed by quantum physics. Anyway, a physical theory speaking about free will and consciousness should also speak about sleep.

According to the proposed quantum model, on the one hand, emotion (``the most basic form consciousness'' \cite{sg}) does not necessarily imply ``the reduction of a quantum state'': in this sense one can have consciousness without reduction. On the other hand, a choice between several possible neuronal assemblies can  very well be described as the jump of a wavefunction into one of its eigenstates. However, the jump may correspond to a conscious choice or an unconscious one. In this sense I assume that human consciousness is not strictly necessary for the collapse of the wave function \cite{kh}. But obviously I also assume that there are conscious collapses and free choices. Overwhelming evidence supporting this assumption, I dare to repeat, comes from the way scientists themselves behave: any publishing scientist would claim to be the conscious and free author of the work he publishes, and not a zombie experiencing hallucinations. Conscious life is more than a dream, even if dreams are important for conscious life.

Finally, the discovery that quantum phenomena involve order beyond time may be relevant for explaining the outcomes of brains: How do neural and cognitive processes correlate to \emph{timeless} mathematical theorems? How is it possible that mortal brains produce so many \emph{immortal} achievements in literature, science, music etc.?

In summary, even if  Quantum Mechanics is currently not being used in neuroscience, one may be confident that quantum effects will appear when the measuring techniques progress.

\section {Conclusions}

If I accept that I am free I have to accept that behind the quantum phenomena there are mighty intellects reading the experimental parameters, calculating the statistical distributions of the events, and deciding the outcomes. Within quantum physics there is place for spiritual agents getting information from the observable world and putting information into it, i.e. guiding the world through the choices they do.

It is possible to unite a deterministic explanation of Gravity with indeterministic Quantum Mechanics, if one accepts that the classical phenomena are also calculated by the same intellects though in a deterministic way. From this point of view, 't Hooft's proposal does not look as ``an important step towards the demystification of quantum mechanics'' but seems rather to inspire the incantation of Classical Mechanics.

The irreversibility of death suggests to use the limit of the human repairing capability for establishing when the ``reduction of the wave function'' takes place. However ``reduction'' can be considered an objective process in the sense that no human observer has to be actually present in order a registration happens. There must be dependence between the measurable physical parameters defining when a detection takes place, and the neuronal assembly size required for consciousness to happen.

Nature calculates the phenomena with the aim of informing human observers. On the one hand, extending this quantum mechanical principle to the theories of Gravity may help to overcome oddities. On the other hand the principle makes clear that the concept of ``the wave function of the entire cosmos'' does not make sense.

In order to describe a world where freedom and creativity are possible it is crucial to develop a neuroscience that overcomes determinism. Quantum Mechanics will surely be of help to this aim.

The ``Quantum Information'' vision that ``information is the stuff the world is made of'', seems to mean after all that the physical reality is made of words non-neuronal intellects speak to neuronal ones.

\end{document}